\documentclass[preprint,aps,tightenlines,floatfix,showpacs]{revtex4}

\usepackage{graphics}
\usepackage{bm}
\usepackage{amsmath}
\usepackage{amssymb}

\begin{document}

\title{Low-energy neutron-$^{12}$C analyzing powers: results from a multichannel 
algebraic scattering theory.}

\author{J. P. Svenne$^{(1)}$}
\email{svenne@physics.umanitoba.ca}
\author{K. Amos$^{(2)}$}
\email{amos@physics.unimelb.edu.au}
\author{S. Karataglidis$^{(2)}$}
\email{kara@physics.unimelb.edu.au}
\author{D. van der Knijff$^{(3)}$}
\email{dirk@unimelb.edu.au}
\author{L. Canton$^{(4)}$}
\email{luciano.canton@pd.infn.it}
\author{G. Pisent$^{(4)}$}
\email{gualtiero.pisent@pd.infn.it}

\affiliation{$^{(1)}$ Department  of  Physics  and Astronomy,
   University  of Manitoba, and Winnipeg  Institute for   Theoretical 
   Physics, Winnipeg, Manitoba, Canada R3T 2N2}
\affiliation{$^{(2)}$ School of Physics, The University of Melbourne, 
   Victoria 3010, Australia}
\affiliation{$^{(3)}$ Advanced  Research   Computing,  Information
   Division, University of Melbourne, Victoria 3010, Australia}
\affiliation{$^{(4)}$ Istituto  Nazionale  di  Fisica  Nucleare,
   Sezione  di Padova, e\\  Dipartimento di Fisica  dell'Universit\`a
   di Padova, via Marzolo 8, Padova I-35131, Italia,}

\date{\today}

\begin{abstract}
  Analyzing powers in low-energy neutron scattering from $^{12}$C are calculated
in an algebraic momentum-space coupled-channel formalism (MCAS).
The results are compared with recently obtained experimental data.
The channel-coupling potentials have been defined previously to reproduce the
total cross section and sub-threshold bound states of the
compound system. Without further adjustment, good agreement with data
for the analyzing powers is obtained.
\end{abstract}


\maketitle

A multi-channel algebraic scattering theory (MCAS) has been formulated
for scattering of nucleons from nuclei, and tested on the well-studied
$^{12}$C nucleus~\cite{Am03}. This formulation of the coupled-channel
scattering theory has the following desirable features: 
(i) the Pauli principle is satisfied even in the context of 
collective nuclear models~\cite{Ca05}; 
(ii) all resonances, no matter how narrow, as well as sub-threshold 
bound states of the compound system are found, without the need of 
calculations on an excessively fine energy mesh; and 
(iii) nuclear structure information can be extracted from the results of the
MCAS calculations, providing the Pauli principle is satisfied~\cite{Pi05}.

Our first work with this formulation focused on calculating cross sections 
for neutron~\cite{Am03} and proton~\cite{Pi05} scattering from $^{12}$C, 
though some sample polarization results have also been shown~\cite{Pi05}. 
Work is in progress on other light nuclear systems, in particular mass 
seven~\cite{Am05a} and mass fifteen~\cite{Ca05b}.
In this paper, we return to $^{12}$C to obtain analyzing powers
which have been recently measured in a detailed study by 
Roper {\it et al}~\cite{Rop05}. The data range in neutron laboratory
(lab.) energy from 2.2 MeV to 8.5 MeV. 
In our MCAS results we limit the energy range to a maximum of 4 MeV.
Beyond that it may be necessary to take account of higher-energy
states of the target nucleus $^{12}$C besides the three we have used so far.

The MCAS formulation yields the complete $S$-matrix for the selected 
scattering system. So, from all entries with the elastic channel, 
and by using standard formulas,
we extract differential cross sections and polarizations as functions of 
the scattering angle and energy. For neutron scattering, the total cross 
section as a function of energy can also be found. 
The total cross section for neutron scattering
from $^{12}$C was published in Refs.~\cite{Am03} and \cite{Pi05}.

The calculations reported here use the same channel-coupling potentials as 
were used in Refs.~\cite{Am03} and \cite {Pi05}; namely, three states of the target nucleus $^{12}$C, the ground ($0^+$) and two excited states 
($2^+$, E = 4.389 MeV, $0_2^+$, E = 7.6542 MeV). 
The coupling to the incoming neutron is via a rotational model potential, 
with quadrupole deformation, $\beta_2 = -0.52$. 
Coupling is taken to second order, and spin-orbit, orbit-orbit and spin-spin 
interactions as well as a central potential are included. 
Closed-shell Pauli blocking effects have been included using 
othogonalizing pseudo-potentials, with very large couplings (1000 MeV).
The potential parameters are those given in Ref.~\cite{Pi05}; 
namely, we have not adjusted any parameters in our
calculation to compare with the recent data.

\begin{figure}[ht]
\scalebox{0.45}{\includegraphics{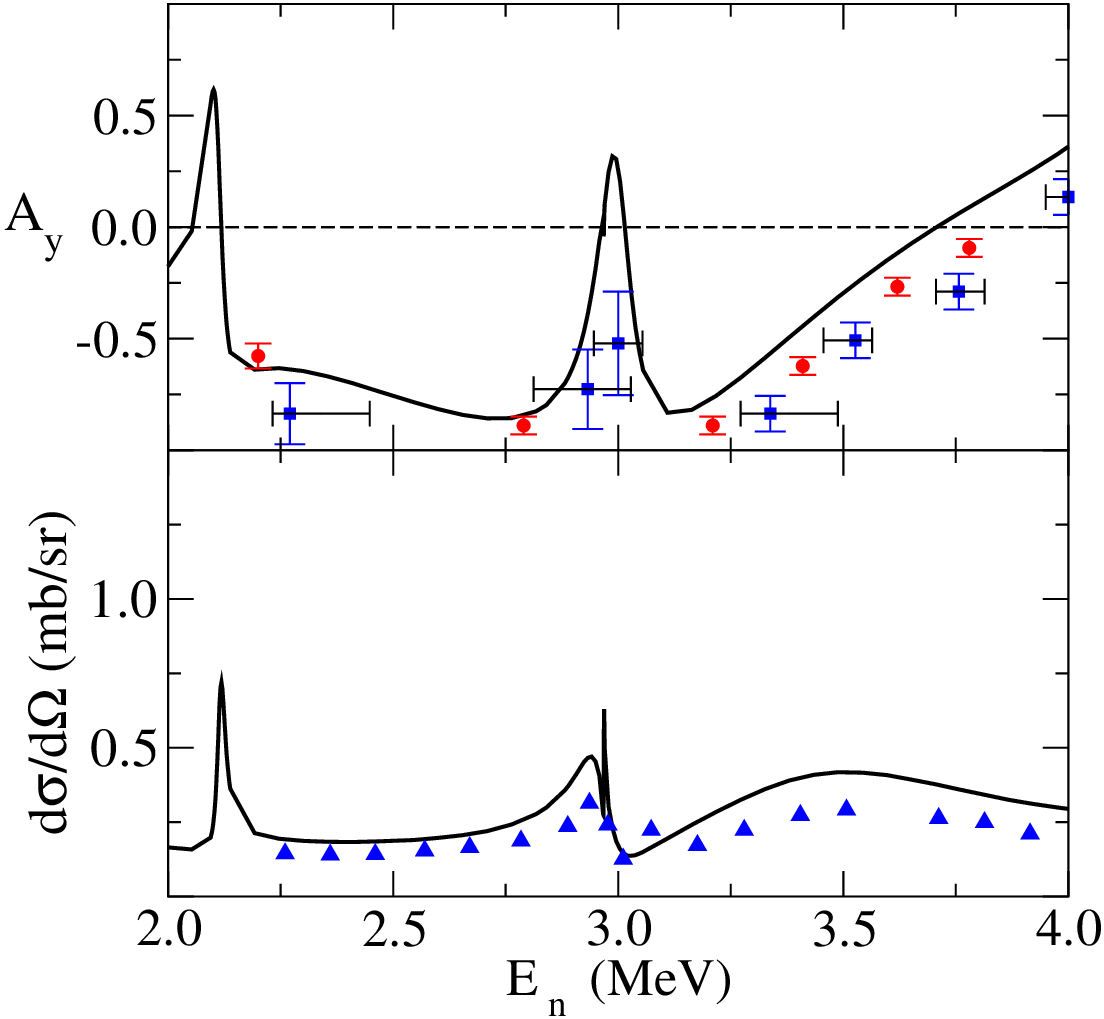}}
\caption{\label{Fig1}(Color online)
The results from MCAS for differential cross section and analyzing
power ($A_y$) at $\theta_{cm} = 43.36^{\circ}$ for the $n+{}^{12}$C system. 
Data (circles) are from Ref.~\cite{Rop05}. 
The last data point is at 41.86$^{\circ}$. The squares with large error bars
represent the 45$^{\circ}$ data from Ref.~\cite{Bu59}. The data for the differential cross section (triangles, lower panel) at 43.03$^{\circ}$ 
are from Ref.~\cite{Fa73}.}
\end{figure}

\begin{figure}[ht]
\scalebox{0.45}{\includegraphics{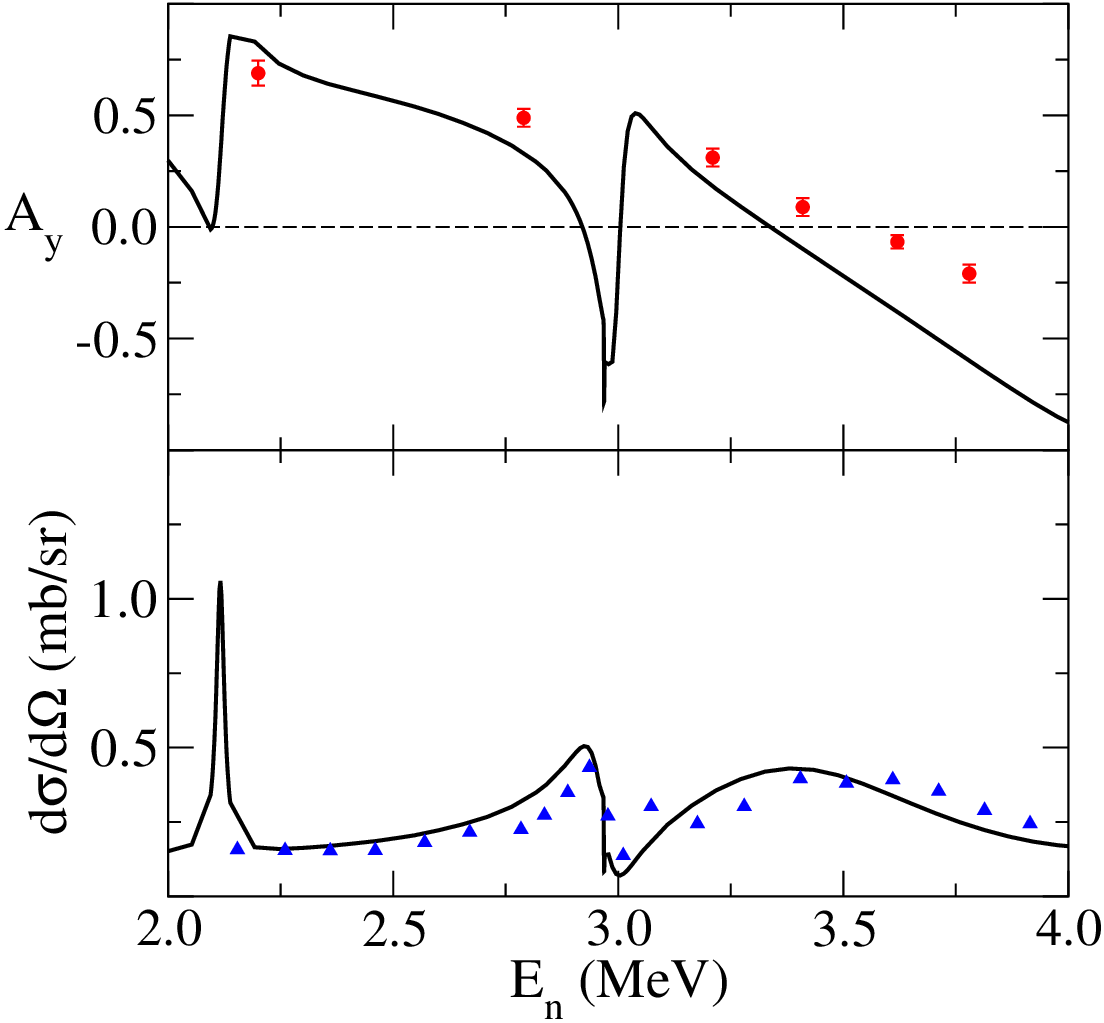}}
\caption{\label{Fig2}(Color online)
The results from MCAS for differential cross section and analyzing
power ($A_y$) at $\theta_{cm} = 147.15^{\circ}$ for the $n+{}^{12}$C 
system. Data (circles) are from Ref.~\cite{Rop05}. 
The last data point is at 146.51$^{\circ}$. The data for the differential
cross section (triangles, lower panel) at 143.03$^{\circ}$
are from Ref.~\cite{Fa73}.}
\end{figure}

In Figs.~\ref{Fig1} and \ref{Fig2}, we show the differential cross section and 
analyzing power as a function of neutron lab. energy ($E_n$)
in the range from 2 to 4 MeV for two center-of-mass
(c.m.)~scattering angles, 43.36$^{\circ}$  and 147.15$^{\circ}$. 
In both cases, the calculations quite satisfactorily match
the data. It should be noted that no adjustment of parameters in the theory 
was done to achieve these results. None of the data points lie on the narrow
resonances at 2.1 and $\sim$ 3.0 MeV, but there is a data point close to the 
first of these. In Fig.~\ref{Fig1}, we show also some older data from 
Bucher {\it et al.}~\cite{Bu59}, taken at 45$^{\circ}$ (square points),
which do have measurements in the region of the $\sim$ 3.0 MeV resonance. 
But the error bars on these data are large.

The MCAS results for differential cross section, solid line in the lower
panel of Figs ~\ref{Fig1} and \ref{Fig2}, have never been published before. 
These theoretical curves compare well with the experimental data from 
Fasoli {\it et al.}~\cite{Fa73}. The error bars in these data are less 
than the size of the triangles indicating the data points. Note that the 
data~\cite{Fa73} are at slightly different angles than the MCAS calculations, 
since those were done at the angles of the $A_y$ measurements 
in Ref.~\cite{Rop05}. 

\begin{figure}[ht]
\scalebox{0.5}{\includegraphics{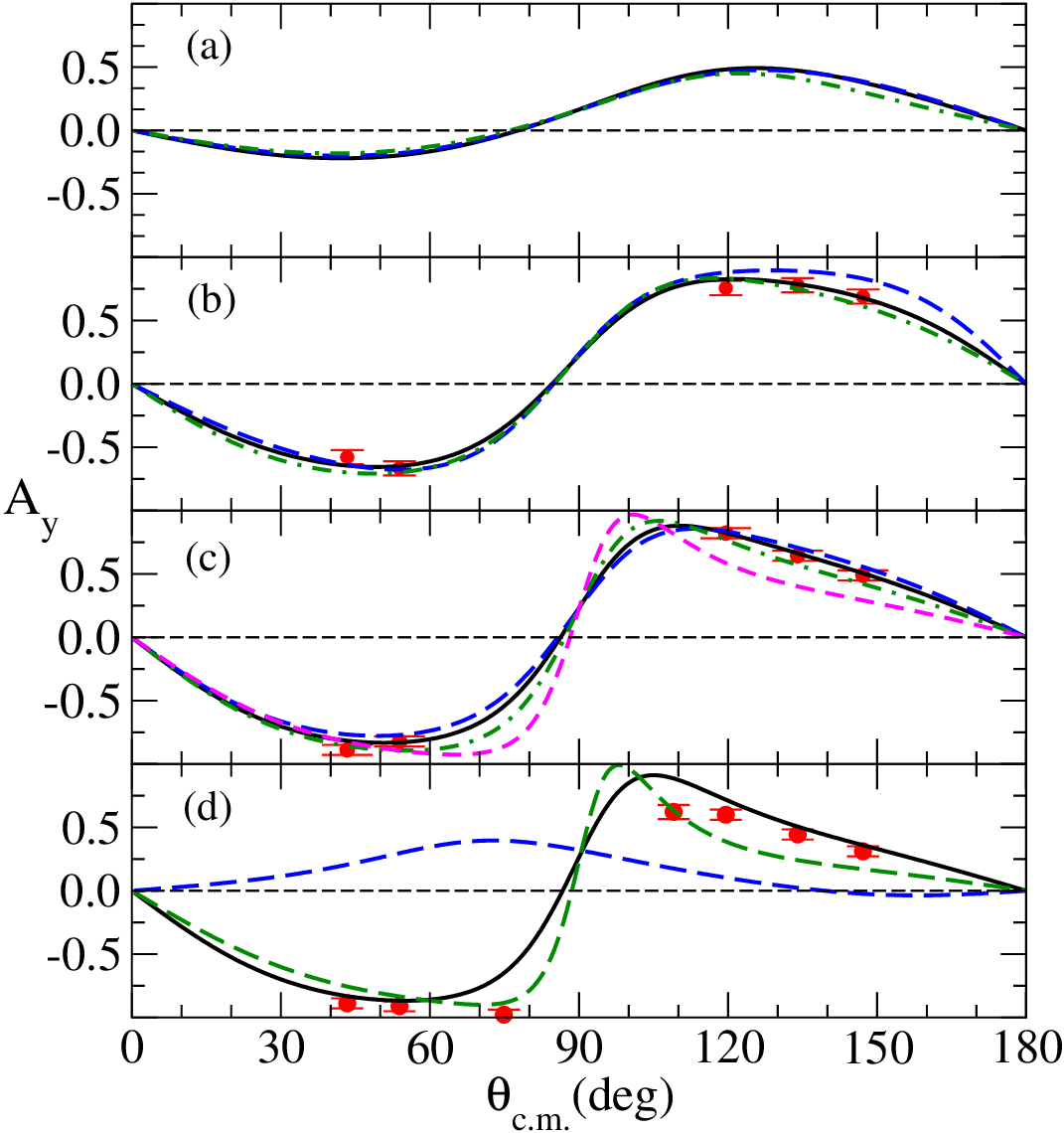}}
\caption{\label{Fig3-5}(Color online)
     The $n+{}^{12}$C analyzing power results from using MCAS theory 
     compared to experimental data from Ref.~\cite{Rop05}. 
Panel (a) represents a prediction of our calculation at 1.9 MeV (solid line).
Dashed line is for 1.8 MeV and dashed-dotted line for 2.0 MeV.
Panel (b): data taken at 2.20 MeV. 
The solid line is the MCAS result for $E_n=$ 2.3 MeV, the dashed/dot-dashed 
lines are for 0.1 MeV below/above that value.
Panel (c): data taken at 2.79 MeV.
The solid line is the MCAS result for $E_n=$ 2.6 MeV, the dashed/dot-dashed
lines are for 0.1 MeV below/above that value. The dotted line is for 2.8 MeV.
Panel (d): data taken at 3.21 MeV.
The solid line is the MCAS result for $E_n=$ 3.1 MeV, the dashed/dot-dashed 
lines are for 0.1 MeV below/above that value.}
\end{figure}

\begin{figure}[ht]
\scalebox{0.5}{\includegraphics{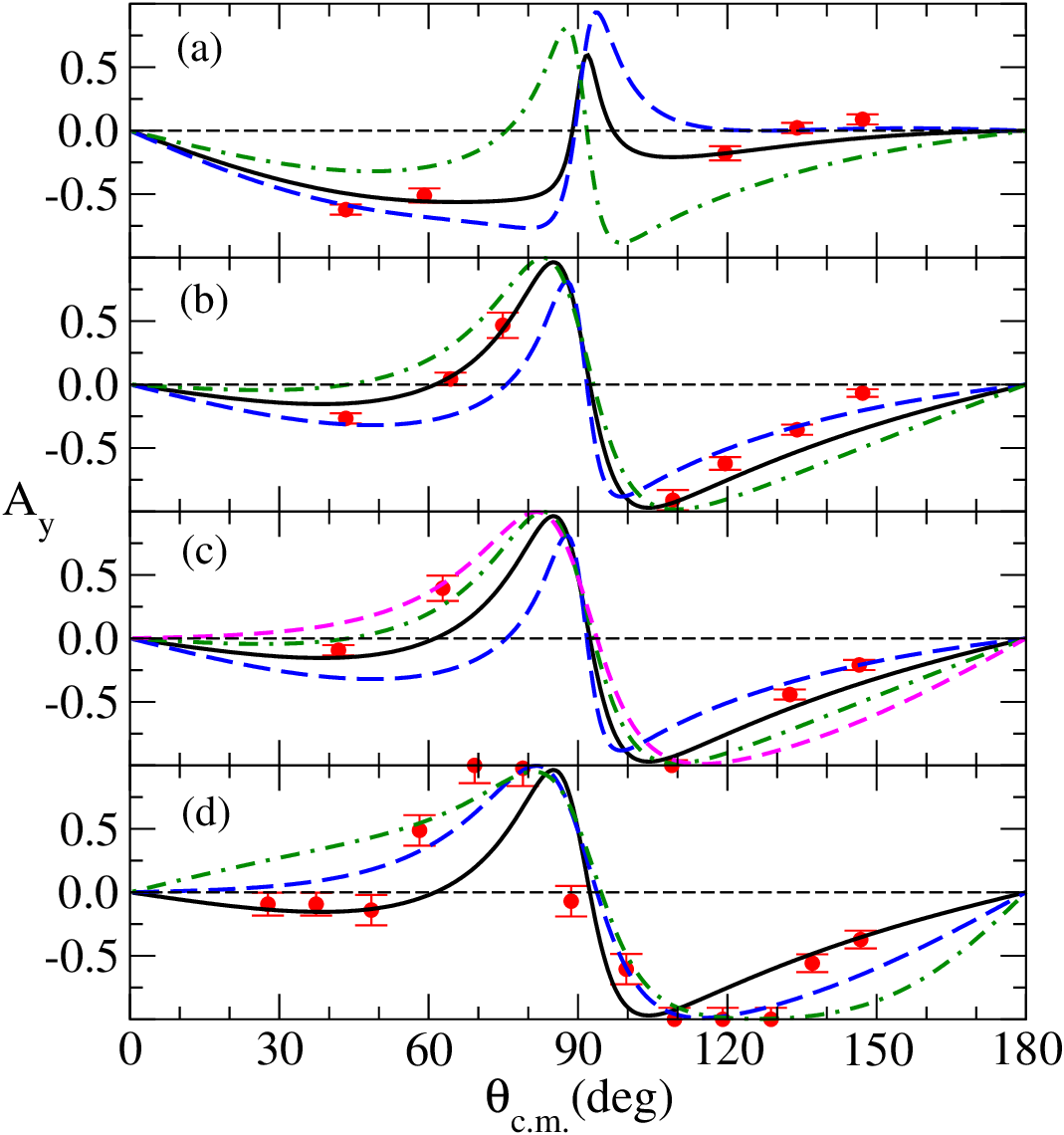}}
\caption{\label{Fig6-8}(Color online)
     The $n+{}^{12}$C analyzing power results from using MCAS theory     
     compared to experimental data from Ref.~\cite{ Rop05}.
Panel (a): data taken at 3.41 MeV.
The solid line is the MCAS result for $E_n=$ 3.4 MeV, the dashed/dot-dashed 
lines are for 0.1 MeV below/above that value.
Panel (b): data taken at 3.62 MeV.
The solid line is the MCAS result for $E_n=$ 3.6 MeV, the dashed/dot-dashed 
lines are for 0.1 MeV below/above that value.
Panel (c): data taken at 3.78 MeV.
The solid line is the MCAS result for $E_n=$ 3.6 MeV, the dashed/dot-dashed
lines are for 0.1 MeV below/above that value. The dotted line is for 3.8 MeV.
Panel (d): data taken at 3.92 Mev. Calculations at 3.6 Mev (solid line), at 
3.8 MeV (dashed line), and 4.0 MeV (dashed/dot line).
}
\end{figure}

\begin{figure}[ht]
\scalebox{0.45}{\includegraphics{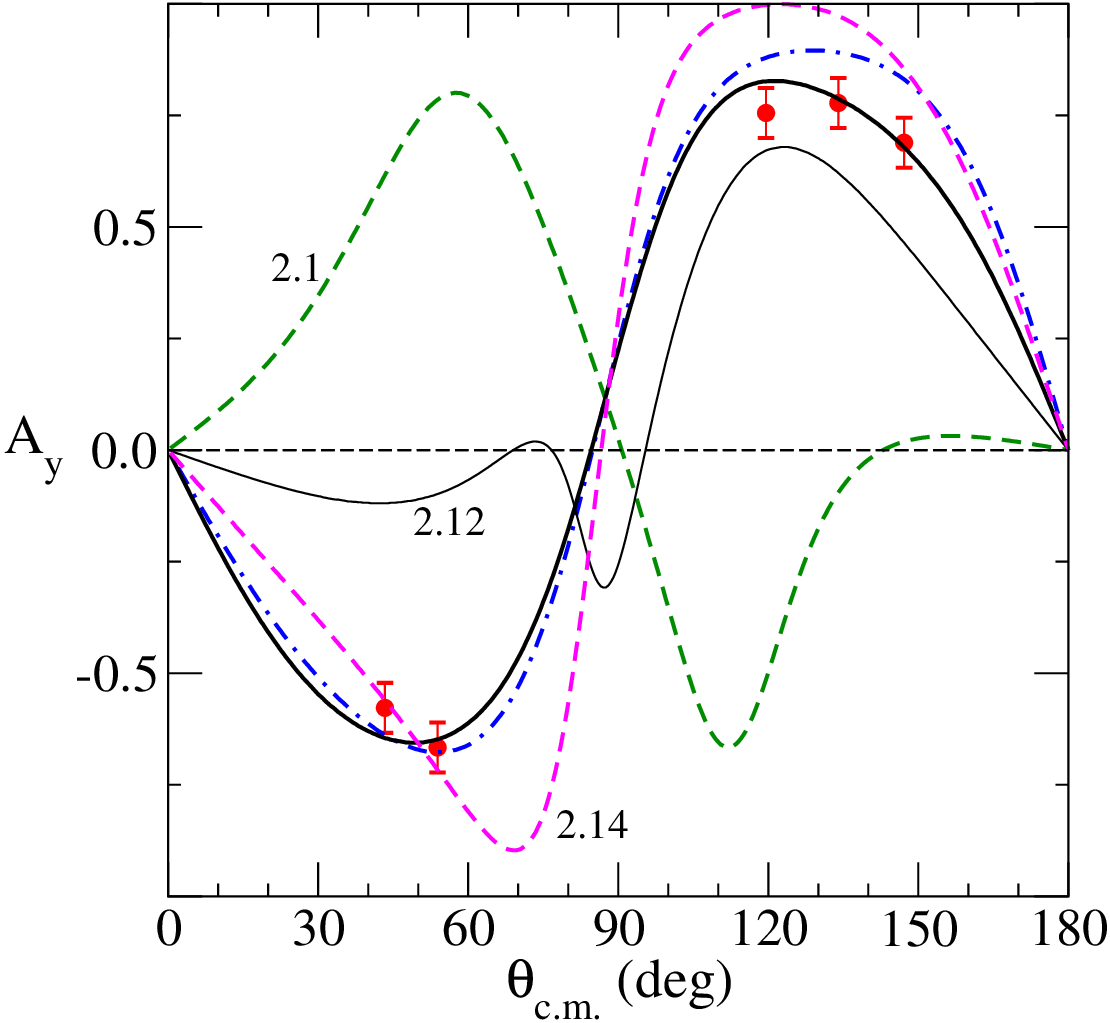}}
\caption{\label{Fig9}(Color online)
     The $n+{}^{12}$C analyzing power results from using MCAS theory,
evaluated at energies between 2.1 and 2.3 MeV  compared to experimental 
data at 2.20 MeV from Ref.~\cite{ Rop05}. The solid line closest 
to the data is for E = 2.3 MeV; the dot-dashed line is for E = 2.2 MeV.
The other three lines have their energies, in MeV, marked next to the line.}
\end{figure}

Next, we show a number of graphs of the analyzing power $A_y$ 
as a function of c.m. angle at various energies in the range 2 to 4 MeV.
In these graphs, Figs.~\ref{Fig3-5} and \ref{Fig6-8}, 
more than one theoretical curve is shown. 
This is because the shape of the analyzing power as a function 
of angle is very sensitive to energy near a resonance. 
Compounding this is the fact that, in the data, 
there is an experimental spread in the neutron lab. energy ranging from 0.2 
to 0.4 MeV, as shown in Fig.~5 of Ref.~\cite{Rop05}. Very near a narrow 
resonance, the sensitivity of shape to energy can be extreme. 
That is most readily seen in our Fig.~\ref{Fig9} which we discuss later.

In Figs. \ref{Fig3-5} and \ref{Fig6-8}, the thick solid line always
represents the theoretical result closest to the data points. 
The long-dashed and dot-dashed lines are for $\pm$0.1 MeV 
different from the energy for the solid line, except for the
highest energy considered, in panel (d) in Fig. \ref{Fig6-8}.
At all but two of the data sets, the theoretical curve agrees with data
within the $\pm$0.2 MeV experimental spread in the neutron energy.
For energy $E_n =$ 2.79 MeV (panel (c) of Fig.~\ref{Fig3-5}) and 
$E_n =$ 3.78 MeV (panel (c) of Fig.~\ref{Fig6-8}), there is a fourth, 
short-dashed, curve at an energy with the result closest to that of the data.
In these two cases the shape of the MCAS result at the quoted experimental 
neutron lab. energy is significantly different from that of the data.
However, it should be noted that, even in these two cases, the difference
between the energy of the closest theoretical representation of the data
and the energy at which the data was taken is just at the limit of the 
experimental uncertainty in the neutron lab. energy. 
Overall, the agreement between the analyzing powers predicted by MCAS 
and the data of Ref.~\cite{Rop05} is very good.

In Fig.~\ref{Fig9} the rapid change with energy of the shape of the 
analyzing power close to the sharp $\left({5 \over 2}\right)^+$ 
resonance at 2.1 MeV is shown.
The curves marked 2.1, 2.12 and 2.14 differ by 20 keV from each other,
but they are very different in shape, and from the shape of the data 
taken at 2.2 MeV.

Determination of the potential parameters (central, spin-spin, 
orbit-orbit, and spin-orbit) was made in 2003
by fitting the overall features of the experimental spectrum
(resonances and bound states)~\cite{Am03}. 
The very good agreement between theoretical
and experimental total cross section (in $n+^{12}$C scattering) demonstrates
that the model correctly describes the process. 
The present analysis on new analyzing power data, Ref.~\cite{Rop05}, 
gives even more confidence in the reliability of the model, since
spin observables are more sensitive quantities to test
the interaction mechanism.
We believe, therefore, that our model can represent a good starting point
for a phase shift analysis.
Important to this is the inclusion of spin-spin and 
orbit-orbit terms, as well as a spin-orbit term
in the interaction potential. Similarily important is the expansion 
to second order in the deformation parameter. 
Most critically needed is the inclusion of the Pauli principle. 
With the MCAS approach, that is done by the use of 
orthogonalizing pseudo-potentials, as detailed in Ref.~\cite{Ca05}.
We conclude that the MCAS theory has predictive power as we 
reproduce data which were not available when the theory was first used to 
fit the total cross section in Ref.~\cite{Am03}.

\begin{acknowledgments}
  This research was supported by a grant from the Australian Research
Council, by a merit award with the Australian Partners for   Advanced
Computing, by the Italian MIUR-PRIN Project  ``Fisica Teorica del
Nucleo e dei Sistemi a Pi\`u Corpi'', and by the Natural Sciences and
Engineering Research Council (NSERC), Canada. KA and JPS also thank
the INFN, sezione di Padova, and the Universit\`a di Padova for financial
support of their visits to Padova for collaboration.

\end{acknowledgments}



\newpage

\newpage
\newpage
\newpage
\newpage

\end{document}